\documentclass[aps,amsfonts,nofootinbib,article,twocolumn,showpacs]{revtex4}
\usepackage{amsmath}
\usepackage{amsfonts}
\usepackage{amssymb}
\usepackage{hyperref}

\def\openone{\leavevmode\hbox{\small1\kern-3.8pt\normalsize1}}
\def\N{\leavevmode\hbox{ Z \kern-8 pt\normalsize{Z}}}
\def\openone{\leavevmode\hbox{\small1\kern-3.8pt\normalsize1}}
\def\openJ{\leavevmode\hbox{J \kern-9.5pt\normalsize J}}
\def\openS{\leavevmode\hbox{ S \kern-9.3pt\normalsize S}}
\newcommand{\bb}{\begin{equation}}
\newcommand{\ee}{\end{equation}}
\newcommand{\eqb}{\begin{eqnarray}}
\newcommand{\eqf}{\end{eqnarray}}

\usepackage{color}

\begin{document}

\title{Construction of Lagrangian and Hamiltonian Structures starting from\\ one Constant of Motion}

\author{Sergio A. Hojman}
\email{sergio.hojman@uai.cl} \affiliation{Departamento de Ciencias,
Facultad de Artes Liberales, Facultad de Ingenier\'{\i}a y Ciencias,
Universidad Adolfo Ib\'a\~nez, Santiago, Chile,\\ and Departamento
de F\'{\i}sica, Facultad de Ciencias, Universidad de Chile,
Santiago, Chile,\\ and Centro de Recursos Educativos Avanzados,
CREA, Santiago, Chile.}

\begin{abstract}
The problem of the construction of Lagrangian and Hamiltonian structures 
starting from two first order equations of motion is presented. This new approach
requires the knowledge of one (time independent) constant of
motion for the dynamical system only. The Hamiltonian and Lagrangian
structures are constructed, the Hamilton--Jacobi equation is then
written and solved and the second (time dependent) constant
of the motion for the problem is explicitly exhibited.

\end{abstract}


\maketitle
\section{Introduction}

Consider a system ${S^1}_{2n}$ of  $2n$ first order differential
equations for $2n$ variables $x^a$,

\begin{equation}
\dot{x}^a = f^a (x^b) ,\ \ \ \ a, b = 1, 2, 3,..., 2n \label{eom}
\end{equation}

A Hamiltonian structure for system ${S^1}_{2n}$ consists of a
Hamiltonian $H$ and a Poisson Bracket relation defined in terms of
an antisymmetric matrix $J^{ab}$ which satisfy the antisymmetry
condition (see, for instance \cite{salmon,sah961,sah962})

\begin{equation}
J^{ab} = - J^{ba}  \label{antis}
\end{equation}

the Jacobi identity

\begin{equation}
J^{ab}_{,d} J^{dc} + J^{bc}_{,d} J^{da} +  J^{ca}_{,d} J^{db} \equiv
0 \label{jacobi}
\end{equation}
and Hamilton equations
\begin{equation}
f^a(x^c)=J^{ab}\frac{\partial H}{\partial x^b}\equiv [x^a,
H]\label{hameq}
\end{equation}

where the Poisson Bracket $[A,B]$ for any pair of dynamical variables
$A(x^a)$ and $B(x^b)$ is defined by

\begin{equation}
[A, B]\equiv \frac{\partial A}{\partial x^a}J^{ab}\frac{\partial
B}{\partial x^b}=-[B, A]\label{pbr}
\end{equation}

A Hamiltonian system ${S^1}_{2n}$ of $2n$ first order differential
equations is said to be (Liouville) integrable if $n$ constants of
motion $C_i,\ (i=1, 2, 3,..., n)$ are known and they are in
involution \cite{liouville,whittaker},

\begin{equation}
[C_i, C_j]=0,\ \forall\ \{i,j\} \label{inv}
\end{equation}

i.e., the Poisson Brackets of all possible pairs of constants
vanish.\\

In order to test Liouville integrability then, the system needs to
be cast in Hamiltonian form with well defined Poisson Bracket
relations.\\

In the usual case, if a Lagrangian for a regular system is known,
then the Hamiltonian structure (canocical momenta, a Hamiltonian and Poisson Bracket
relations) may be constructed using standard
textbook procedures (see, \cite{goldstein}, for instance).\\

Nevertheless, if a Lagrangian is not known (or if it fails to exist)
the usual textbook procedure is no longer applicable.\\

Techniques to construct a Hamiltonian structure have been devised
and they require the knowledge of at least one constant of the
motion and symmetry vectors of the differential system \cite{sah961,sah962}.\\

In this article we present new results regarding the case $n=1$
where one constant of motion allows for the construction of the
Hamiltonian and Lagrangian structures as well as the complete
integration of the problem. Solutions obtained previously require the knowledge 
of two constants of motion \cite{darboux,cs,hh,hu} or a constant of motion 
and a symmetry vector \cite{sah961,sah962}.\\

The new procedure is applied to motion of projectiles subject to air
drag as well as to other dynamical equations.\\

\section{Construction of a Hamiltonian Structure}

Consider a system ${S^1}_{2}$ of  two first order differential
equations for two variables $x^a$ \cite{darboux,cs,hh,hu},

\begin{equation}
\dot{x}^a = f^a (x^b).\ \ \ \ a, b = 1, 2 \label{eom2}
\end{equation}

Assume one time independent constant of motion $C_1=C_1 (x^b)$ is
known. The (time independent) constant of motion $C_1(x^b)$ satisfies

\begin{equation}
\frac{\partial C_1}{\partial x^a} f^a (x^b) \equiv 0.\ \ \ \ a, b =
1, 2 \label{cm}
\end{equation}

The construction of a Hamiltonian structure for system \eqref{eom2}
requires the knowledge of a Hamiltonian $H$ and a Poisson Bracket
antisymmetric matrix $J^{ab}$. Nevertheless, in dimension $2$ there
is essentially one antisymmetric matrix, therefore,

\begin{equation}
J^{ab}= \left( \begin{array}{ccc}
0 & \mu (x^b)  \\
-\mu (x^b) & 0  \\
\end{array} \right),
\label{2dpb}
\end{equation}

where the function $\mu (x^b)$ is determined by the Hamilton
equations condition
\begin{equation}
f^a(x^c)=J^{ab}\frac{\partial C_1}{\partial x^b},\label{2dhameq}
\end{equation}

where the choice $H=C_1$ has been made. Note that due to \eqref{cm}
the gradient of $C_1$ is orthogonal to $f$ so that
$J^{ab}\frac{\partial C_1}{\partial x^b}$ is parallel to $f^a$ in a
two dimensional space, thus condition \eqref{2dhameq} determines the
function $\mu(x^b)$ uniquely.\\

It is worth mentioning that in a two dimensional space the Jacobi
identity \eqref{jacobi} is always satisfied by any antisymmetric
matrix.\\

Therefore, if a time independent constant of motion $C_1(x^b)$ for
system \eqref{eom2} is known, the Hamiltonian structure is defined
by choosing the Hamiltonian $H=C_1$ and the Jacobi matrix $J^{ab}$
is completely determined by the antisymmetry requirement
\eqref{2dpb} and the Hamilton equations condition \eqref{2dhameq}.\\ 

Furthermore, Liouville integrability criterion is always met with one constant 
of motion in a two dimensional phase space.\

\section{Construction of a Lagrangian Structure}

Hamilton equations \eqref{2dhameq} may be rewritten as

\begin{equation}
\dot{x}^a = J^{ab}\frac{\partial C_1}{\partial x^b}\label{2dhameq2}
\end{equation}

Introduce the
antisymmetric Lagrange Brackets matrix ${\sigma}_{ab}= -{\sigma}_{ba}$

\begin{equation}
J^{ab}{\sigma}_{bc}=-{{\delta}^a}_c \label{Lag}
\end{equation}

which is (up to a sign)  the matrix inverse of the Poisson Brackets matrix $J^{ab}$ .
Left multiply \eqref{2dhameq2} by ${\sigma}_{ca}$, to get the Lagrangian form of Hamilton equations,

\begin{equation}
{\sigma}_{ca} \dot{x}^a +\frac{\partial C_1}{\partial x^c}=0. \label{ELeq}
\end{equation}

In matrix form ${\sigma}_{ab}$ is
\begin{equation}
{\sigma}_{ab}= \left( \begin{array}{ccc}
0 & \frac{1}{\mu (x^b)}  \\
-\frac{1}{\mu (x^b)} & 0  \\
\end{array} \right).
\label{2dlb}
\end{equation}

Consider the Lagrangian $L = L({x}^a, \dot{x}^b)$

\begin{equation}
L = L({x}^a, \dot{x}^b)= l_1 ({x}^a) \dot{x}^1 - C_1 ({x}^a),\label{Lag1}
\end{equation}
where $l_1 ({x}^a)$ is defined by

\begin{equation}
\frac{\partial l_1}{\partial {x}^2} = \frac{1}{\mu}. \label{Lag2}
\end{equation}

The Euler Lagrange equations for \eqref{Lag1} are
\begin{equation}
\frac{\partial l_1}{\partial {x}^2}\dot{x}^2 + \frac{\partial
C_1}{\partial {x}^1}=0 \label{eom2}
\end{equation}

and
\begin{equation}
-\frac{\partial l_1}{\partial {x}^2}\dot{x}^1 + \frac{\partial
C_1}{\partial {x}^2}=0 \label{eom3}
\end{equation}
which are identical to \eqref{ELeq} and equivalent to
\eqref{2dhameq2} once \eqref{2dpb} and
\eqref{Lag2} are taken into account.\\

Note that $l_1 ({x}^a)$ is determined only up to the addition of an
arbitrary function $f_1 ({x}^1)$. This addition modifies the Lagrangian
\eqref{Lag1} by a total time derivative, which means that equations
\eqref{eom2} and \eqref{eom3} remain invariant under the change.\\

Now define phase space variables $q$ and $p$ by

\begin{equation}
q\equiv x^1, \label{q}
\end{equation}

and

\begin{equation}
p\equiv l_1 ({x}^a). \label{p}
\end{equation}

It is a straightforward matter to compute the Poisson Bracket

\begin{equation}
[q, p] = 1, \label{qp}
\end{equation}

using \eqref{pbr}, \eqref{2dpb} and  \eqref{Lag2}, to realize that $q$ and $p$ are canonically conjugated variables.
Note that the non uniqueness in the definition of $l_1 ({x}^a)$ generates canonical transformations in $p$ leaving $q$ invariant.

\section{Construction of the Hamilton--Jacobi equation and the general solution of the system}

Consider now the inverse transformation of \eqref{q}, \eqref{p} given by

\begin{equation}
x^1 = q, \label{x1}
\end{equation}

and

\begin{equation}
x^2 = g(q,p), \label{x2}
\end{equation}

where the last equation is obtained by solving \eqref{p} for $x^2$.
We have used the fact that

\begin{equation}
\frac{\partial l_1}{\partial x^2} \neq 0, \label{l1x2}
\end{equation}
so that one can always solve \eqref{p} for $x^2$.\\

Define now the Hamiltonian ${H(q,p)}$ by

\begin{equation}
H(q,p) = C_1(q,g(q,p)), \label{H}
\end{equation}

Therefore, the Hamilton--Jacobi equation is

\begin{equation}
H(q,\frac{\partial S}{\partial q})+ \frac{\partial S}{\partial t}= 0, \label{HJ1}
\end{equation}

or

\begin{equation}
H(q,\frac{\partial W}{\partial q})= C_1, \label{HJW}
\end{equation}

where

\begin{equation}
S (q, C_1, t) = W (q, C_1) - C_1 t \label{HJS1}
\end{equation}

The second (explicitly time dependent) constant of motion $C_2$ for the system
is obtained by solving \eqref{HJW} for $W (q, C_1)$, (i.e., solving
for $\frac{\partial W}{\partial q}$ and integrating) and defining

\begin{equation}
C_2= \frac{\partial S}{\partial C_1}= \frac{\partial W}{\partial C_1} - t,  \label{HJS2}
\end{equation}

thus one gets the general solution to the problem (up to computing an integral).

\section{Examples}
\subsection{One second order equation}

Consider the dynamics of a system defined by

\begin{equation}
\ddot q= F(q)G(\dot q) \label{soeom}
\end{equation}

One second order equation may, of course, be written as a two
dimensional first order system.\\

Define
\begin{equation}
x^1 \equiv q, \label{def1}
\end{equation}
and
\begin{equation}
x^2 \equiv \dot q. \label{def2}
\end{equation}

The equations of motion
\begin{equation}
{\dot x}^1= x^2 \label{soeom1}
\end {equation}
and
\begin{equation}
{\dot x}^2= F(x^1)G(x^2) \label{soeom2}
\end {equation}
are equivalent to \eqref{soeom} and definitions \eqref{def1} and
\eqref{def2}.\\

A time independent constant of motion $C_1(q, \dot q)$ for
\eqref{soeom} is given by

\begin{equation}
C_1(q, \dot q) = - \int {F(q) dq} +\int {\frac{\dot q}{G(\dot q)}
d\dot q}. \label{c1}
\end{equation}

Therefore, the Hamiltonian $H$ is

\begin{equation}
H(x^1, x^2) = - \int {F(x^1) dx^1} +\int {\frac{x^2}{G(x^2)} d x^2},
\label{Hex1}
\end{equation}

and the Poisson Bracket matrix can be written as

\begin{equation}
J^{ab}= \left( \begin{array}{ccc}
0 & G(x^2)  \\
- G(x^2) & 0  \\
\end{array} \right)
\label{pbex1}
\end{equation}
to reproduce \eqref{soeom1} and \eqref{soeom2}.\\

The momentum $p$ is given by

\begin{equation}
p=\int \frac{dx^2}{G(x^2)}  \label{pex1}
\end{equation}

One can now proceed following the steps described in Section IV. To
be more concrete, consider a few explicit functional forms for
$G(\dot q)$

\subsubsection{$G(\dot q)=1$}

In this case (as a matter of fact, for any constant $G(\dot q)$),
one gets that $C_1$ is the energy, $p=x^2$ and the Hamilton--Jacobi
equation is

\begin{equation}
\frac{1}{2}\left({\frac{\partial S}{\partial q}}\right)^2-\int F(q)
dq + \frac{\partial S}{\partial t}= 0, \label{HJex11}
\end{equation}
where

\begin{equation}
S(q, C_1, t) = W(q, C_1)- C_1 t, \label{Sex1}
\end{equation}
and $W(q, C_1)$ satisfies

\begin{equation}
\frac{1}{2}\left({\frac{\partial W}{\partial q}}\right)^2-\int F(q)
dq = C_1. \label{HJ1ex11}
\end{equation}

The second (time dependent) constant of motion $C_2$ is given by

\begin{equation}
C_2= \frac{\partial S(q, C_1, t)}{\partial C_1} = -t + \int
\frac{dq}{{\sqrt{2(C_1+\int F(q') dq')}}}, \label{C2ex1}
\end{equation}
which completes the solution.

\subsubsection{$G(\dot q)=\dot q$}

Now

\begin{equation}
C_1= - \int{F(q) dq} +\dot q \label{c1ex2},
\end{equation}
and
\begin{equation}
p= \int\frac {d\dot q }{\dot q } = \ln \dot q  \label{pex2}.
\end{equation}

The Hamitonian is
\begin{equation}
H(q, p)= - \int{F(q) dq} +e^p \label{Hex2},
\end{equation}

and the Hamilton--Jacobi equation is 
\begin{equation}
 e^\frac{\partial S}{\partial q}- \int{F(q) dq} + \frac{\partial S}{\partial t}= 0. \label{HJex2}
\end{equation}

The solution to the Hamilton--Jacobi equation is
\begin{equation}
S(q, C_1, t) = \int dq\ {\ln\left(C_1+ \int{F(q') dq'}\right)}- C_1 t, \label{Sex2}
\end{equation}
The second constant $C_2$ is

\begin{equation}
C_2= \frac{\partial S(q, C_1, t)}{\partial C_1} = -t + \int
\frac{dq}{{(C_1+\int F(q') dq')}} \label{C2ex2}
\end{equation}
which completes the solution.

\subsection{Motion of projectiles subject to drag forces}

The trajectories of projectiles moving on a medium in which drag
forces are not negligible are of potential practical importance to
ballistics and sports, for instance, in addition to their interest
associated to theoretical problems related to solving ordinary
differential equations and constructing Lagrangian and Hamiltonian
structures as well as the associated Hamilton--Jacobi equation.\\

Parabolic motion is the well known solution for the trajectory of a
projectile moving on the surface of the Earth neglecting air
resistance.\\

The actual motion in the presence of a viscous medium is, of course,
different when drag forces are important. Parker dealt with such a
problem in an article published in 1977 \cite{parker} where the general solution
as well as applications to different regimes of projectile
motion were presented. Recently, a different approach to solve the same problem 
was presented by Shouryya Ray, an Indian born german highschool student
but there does not seem to be any published record of his findings. Nevertheless, see
http://bit.ly/KLgGYd  \\

Some time ago, it was commonplace to state that dissipative problems
were out of the realm of Lagrangian and/or Hamiltonian
descriptions.\\

Nevertheless, in the last decades, different procedures to relate
symmetries and conserved quantities to the construction of
Lagrangian and Hamiltonian structures have been devised 
\cite{sah961,sah962,cgr1,cgr2}.\\

Consider a projectile moving near the surface of the Earth subject to
gravity and air drag force proportional to the square of its speed.
Let $u$ and $v$ be the horizontal and vertical components of its
velocity, respectively. The equations of motion are 
(following http://bit.ly/KLgGYd)

\begin{equation}
 \dot{u}= - a u\sqrt{u^2+v^2},\
\label{equ}
and
\end{equation}
\begin{equation}
 \dot{v}= - a v\sqrt{u^2+v^2}-g,\
 \label{eqv}
\end{equation}

where $a$ and $g$ are constants.\\

Multiply (\ref{equ}) times $v$ and (\ref{eqv}) times $u$ and subtract to get
\begin{equation}
  v \dot{u}-  u \dot{v} = g u
\label{equv}
\end{equation}
or
\begin{equation}
  \frac{d}{dt}({\frac{v}{u}}) = - \frac{g}{u}
\label{eqqp}
\end{equation}
Now multiply (\ref{equ}) times $\dot{v}$ and (\ref{eqv}) times
$\dot{u}$ and subtract to get
\begin{equation}
  -a( u \dot{v}-  v \dot{u}) \sqrt{u^2+v^2}+ g \dot{u}=0
\label{equv}
\end{equation}

or
\begin{equation}
  a \ \sqrt{1+({\frac{v}{u}})^2}\ \frac{d}{dt}({\frac{v}{u}})- g\ \frac{\dot{u}}{u^3}=0
\label{equv}
\end{equation}
Define
\begin{equation}
q_1\equiv {\frac{v}{u}} \label{qdef}
\end{equation}
and
\begin{equation}
q_2\equiv - {\frac{1}{u}} \label{pdef}
\end{equation}

to get
\begin{equation}
 \dot{q_1}= g\ q_2
\label{eqq}
\end{equation}

and

\begin{equation}
a \ \sqrt{1+{q_1}^2}\ \dot{q_1}+\ \frac{1}{2}\ g\
\frac{d({q_2}^2)}{dt}  =\ 0 \label{eqp}
\end{equation}

Equation (\ref{eqp}) may be readily integrated to yield a constant
of motion $C_1$
\begin{equation}
C_1 =\frac{1}{2}  \left[ a \left(q_1 \sqrt{1+{q_1}^2}+ {sinh}^{-1}
(q_1)\right)
  +\ g\ {q_2}^2 \right].
  \label{H}
\end{equation}
As a matter of fact, a Hamiltonian structure defined by canonical
variables $q\equiv q_1$, $p\equiv q_2$ and Hamiltonian $C_1(q,p)$ is such
that Hamilton equations for this system are equivalent to the
equations of motion (\ref{eqq}) and (\ref{eqp}) as it can be
straightforwardly realized. In fact, one Lagrangian $L$ for equations \eqref{eqq} and \eqref{eqp} may be written as

\begin{equation} 
L= p\ \dot q - H (q, p) \label{airdragL}
\end{equation} 

The Hamilton--Jacobi equation for such a system is

\begin{equation}
\frac{1}{2}  \left[ \ g\ \left({\frac{\partial S}{\partial
q}}\right)^2
  + a \left(q \sqrt{1+{q}^2}+ {sinh}^{-1} (q)\right)\right]+ \frac{\partial S}{\partial t} =0.
  \label{HJ1}
\end{equation}

Therefore, as usual,
\begin{equation}
S(q,E,t)= W(q,E)-Et \label{S}
\end{equation}

where
\begin{equation}
\frac{1}{2}  \left[ \ g\ \left({\frac{\partial W}{\partial
q}}\right)^2
  + a \left(q \sqrt{1+{q}^2}+ {sinh}^{-1} (q)\right)\right]=E.
  \label{HJ2}
\end{equation}

Solve for $W$ to get

\begin{equation}
W = \int dq \sqrt{\frac{\left(2E-a \left(q \sqrt{1+{q}^2}+
{sinh}^{-1}(q)\right)\right)}{g}}
  \label{W1}
\end{equation}
and the time dependent constant of motion $t_0$ is

\begin{equation}
t_0= -\frac{\partial S}{\partial E}
  \label{t01}
\end{equation}

\begin{equation}
t_0= t-\int \frac{dq}{\sqrt{\left(2gE-ag \left(q \sqrt{1+{q}^2}+
{sinh}^{-1}(q)\right)\right)}},
  \label{t02}
\end{equation}
which completes the solution of the problem.\\

\section{Conclusions}

The usual approach to Lagrangian and Hamiltonian dynamics assumes the knowledge of 
a Lagrangian from which the (Euler--Lagrange) equations of motion, 
the canonical momenta, the Poisson Brackets relations, the Hamiltonian, Hamilton's and Hamilton--Jacobi 
equations are derived (and sometimes solved). One tool to test integrability of the dynamical 
differential equations is Liouville's theorem based on the structures described above. 
The quantization of the classical system may also be achieved using these exact same structures.\\

In other words, the Lagrangian is an extremely powerful tool which allows us to construct 
all the dynamical entities which are needed to study a classical dynamical system and to quantize it.\\

In this article a different approach is presented. The building blocks are the equations of motion 
(no prior knowledge of a Lagrangian or a Hamiltonian structure is assumed) and one time independent constant 
of motion for a two dimensional first order system.\\

These ingredients are enough to construct a Hamiltonian, Poisson Brackets relations, a Lagrangian,
a canonical momentum,  Hamilton's and Hamilton--Jacobi equations as well as the second
constant of the motion of the problem, therefore solving the Inverse Problem of the Calculus of Variations 
and getting the general solution to the differential system.\\

Previous approaches to the solution of these problems require prior knowledge of two constants 
of motion \cite{darboux}, \cite{hu} or of a constant of motion and a symmetry vector \cite{sah961}, which amounts to knowing the general solution of the problem.\\
 
Therefore, the approach presented here represents a real progress compared to previous methods 
in the sense that less stringent requirements are needed and the solution to the problem is explicitly provided (instead of required).

\end{document}